%% file: skeleton.tex
\documentclass[a4paper,11pt]{article}
\usepackage{pos}
\usepackage{hyperref}
\usepackage{lineno}
\usepackage{wrapfig}
\usepackage{graphicx}
\usepackage{capt-of}   
\title{Determining the orientation of radio antennas at the South Pole using Galactic noise measurements }
\ShortTitle{Estimating antenna orientation using Galactic Noise}

\author{The IceCube Collaboration \\{\normalsize \normalfont(a complete list of authors can be found at the end of the proceedings)}\\}

\emailAdd{paras.koundal@icecube.wisc.edu}

\abstract{The IceCube Neutrino Observatory is a multi-messenger observatory at the South Pole. As preparation for an enhancement of its surface array, IceTop, a prototype station consisting of elevated scintillation panels and radio antennas has been installed and is operating since 2020. The radio antennas detect emissions from cosmic-ray-induced air showers, and their precise orientation is essential for an accurate reconstruction of the air-shower properties. This work presents a novel method to determine the orientation by analyzing periodic variations of the Galactic background noise recorded by the antennas. In particular, we examine noise level variations correlated with the Earth's rotation and the apparent position of the Galactic Center. The method can provide a potential alternative or augment GPS-based measurements of the alignment of radio antennas at the South Pole.

\vspace{4mm}
{\bfseries Corresponding authors:}
Paras Koundal$^{1*}$,
Valeria Torres-Gomez$^{2}$,\\
{$^{1}$ \itshape Bartol Research Institute, Department of Physics and Astronomy, University of Delaware, U.S.A.}\\
{$^{2}$ \itshape Universidad de los Andes, Colombia}\\[4mm]
$^*$ Presenter}

\ConferenceLogo{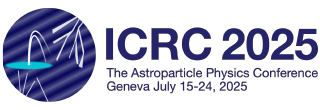}

\FullConference{39th International Cosmic Ray Conference (ICRC2025)\\
 15–24 July 2025\\
Geneva, Switzerland\\}

\newcommand{\PZero}{Pol.-0~}
\newcommand{\POne}{Pol.-1~}

\begin{document}
\maketitle

\section{Introduction}\label{sec:intro}
The IceCube Observatory, the megaton-scale ice Cherenkov detector located close to the geographic South Pole, is a powerful neutrino and cosmic-ray observatory. IceTop (IT) which is the surface component of the observatory detects the electromagnetic (EM) and muonic component of extensive air showers (EAS).  Since the deployment completion in late 2011, the mean snow accumulation over all the IT tanks has increased. The increasing and non-uniform snow accumulation over tanks leads to uncertainty in signal-strength expectation at IT and an increasing detection threshold. Even though ongoing works \cite{Rawlins2023} are developing techniques to improve the model for snow attenuation, it still remains a major systematic for most of the IT analysis. To overcome this, surface array enhancement (SAE) including elevated scintillation detectors and radio-antennas was proposed \cite{Haungs2019}. Since January 2020, a prototype station of the planned SAE consisting of 8 scintillators and three radio antennas (triggered by scintillators) has been collecting data at South Pole. In January 2025, two more stations have been deployed at South Pole \cite{ShefaliICRC2025, MeghaICRC2025}. SAE is planned to be integrated into the Surface Array of the planned next-generation of the observatory, namely IceCube-Gen2 \cite{Aartsen2021}.\par 
\begin{wrapfigure}{r}{0.4\textwidth}
    \centering
    \includegraphics[width=\linewidth]{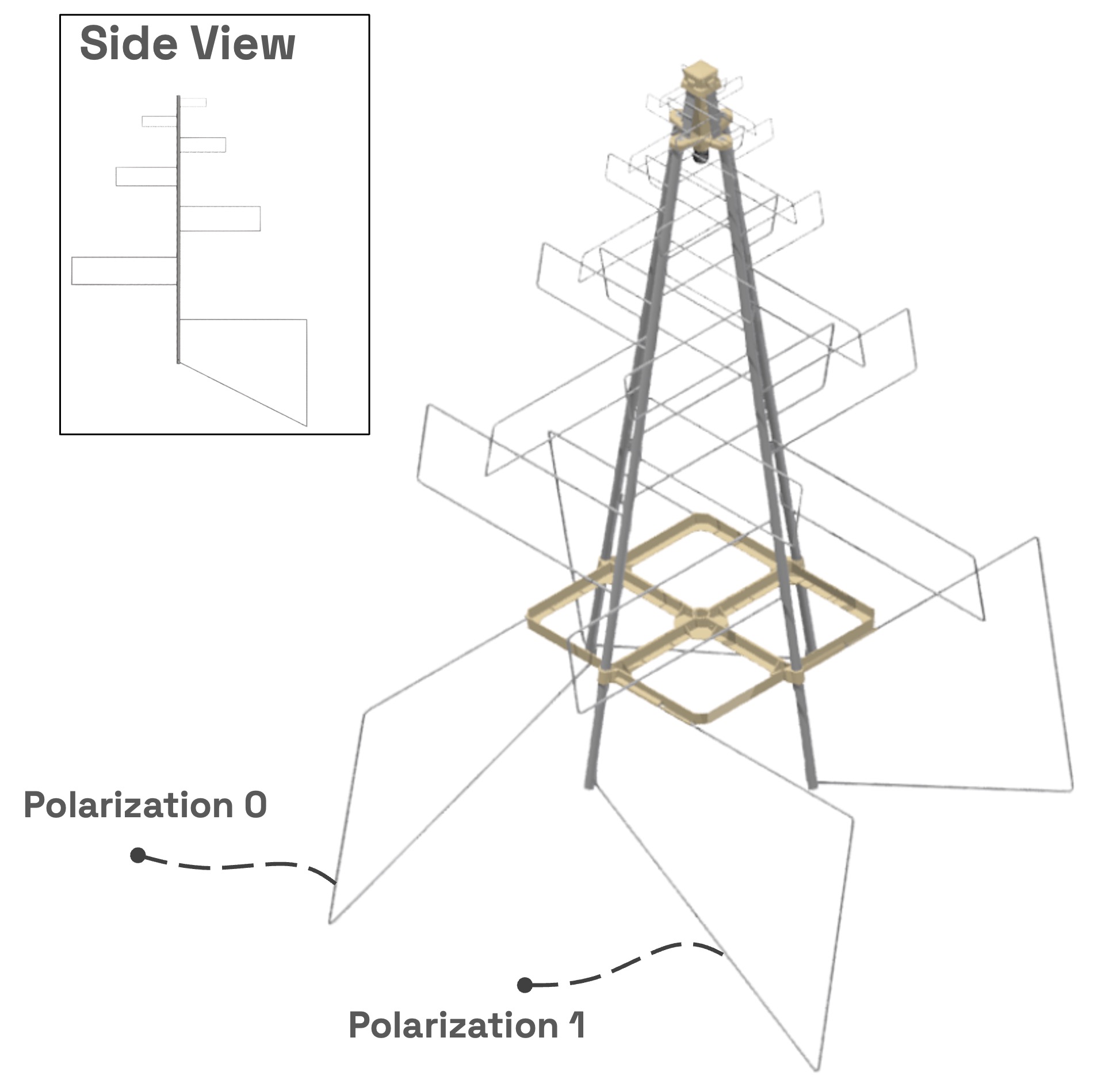}
    \caption{3D model of SKALA-v2 radio-antenna used in the prototype station at IceCube.}
    \label{fig:antenna}
\end{wrapfigure}

This work utilizes waveforms measured with the radio antennas of the SAE prototype station in 2023. A three-dimensional model of the antenna used for the measurements is shown in \autoref{fig:antenna}. Each antenna consists of two perpendicular polarizations/channels, recording waveforms independently, labeled as Polarization 0 (or \PZero) and Polarization 1 (or \POne). This work aims to use the time variation of the strength of measured galactic background noise originating from the Milky Way \cite{refId0} in order to determine the rotational orientation of the Surface Array antennas at the IceCube Neutrino Observatory. Traditionally, antenna orientations within the array have been measured via differential GPS surveys, which involve measuring the positions of the endpoints of each antenna arm. The most recent GPS measurement for the SAE was conducted in December 2024 and will be used for subsequent comparisons of antenna orientations.  While effective, this method is resource-intensive, requires specialized equipment, and can only be conducted during rare field campaigns. In contrast, our approach seeks to exploit the constant visibility of the galactic-center (GC) from the antennas. It is visible at nearly constant zenith angle at the South Pole and performs one complete revolution around the geographic South Pole during a sidereal day. Hence, the GC can serve as a natural and persistent radio source for antenna alignment. By analyzing how the measured background noise varies as the antenna orientation changes relative to the GC, we aim to remotely infer the antenna orientation based on radio data. The key hypothesis assumed in this analysis is that the maxima in the galactic noise signal occur when the antenna arm is oriented perpendicular to the direction of the signal primarily coming from the GC. Hence, by utilizing the GC direction in the sky, we can determine the antenna alignment without relying on GPS-based surveys.\par
Understanding the orientation of an antenna is crucial since the sensitivity of a radio antenna has significant angular dependence i.e. input signal of the same strength will have different sensitivity of detection based on the direction of the source w.r.t. antenna orientation direction. The data driven method shows promise in potentially augmenting the GPS measurements on yearly basis. This study hence presents a new way to use galactic noise, previously applied only to calibrate the gain of radio antennas \cite{MULREY20191}. In the following text, \autoref{sec:method} will discuss the details of cleaning the radio-waveforms, calculating the galactic-signal strength and outlier removal. This is utilized to get monthly estimates of the antenna orientation. The monthly estimates are utilized to get yearly estimate of the antenna orientations and is compared to the GPS measurements in \autoref{sec:results}. Finally, \autoref{sec:outllok} will discuss the summary and outlook of the work. 
\section{Methodology}\label{sec:method}
\begin{figure}[t]
  \centering
  \makebox[\textwidth][c]{%
    \begin{minipage}{0.65\textwidth}
      \includegraphics[width=\linewidth]{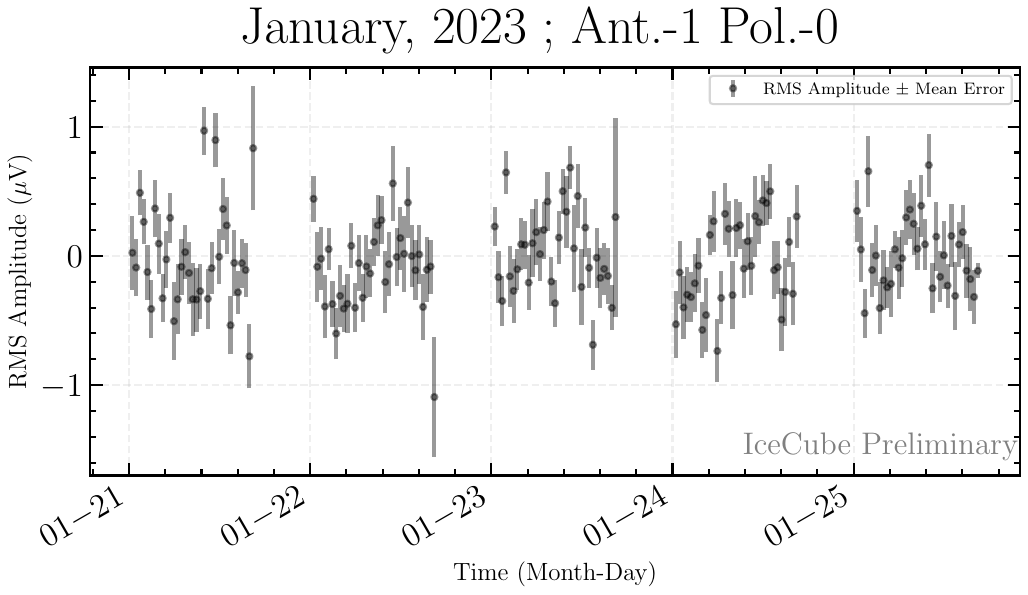}
    \end{minipage}\hfill
    \begin{minipage}{0.32\textwidth}
      \captionof{figure}{%
        Time variation of RMS values over several days in January 2023, after applying initial cleaning and the 2D RMS-based cut. The emerging oscillatory behavior in RMS is consistent with expectations from the rotation of the galactic center relative to the antenna array. Gaps occur because measurements aren't performed throughout the entire day.
      }\label{fig:timeVar}
    \end{minipage}%
  }
\end{figure}

\begin{figure}[t]
  \centering
  \makebox[\textwidth][c]{%
    \begin{minipage}{0.55\textwidth}
      \includegraphics[width=\linewidth]{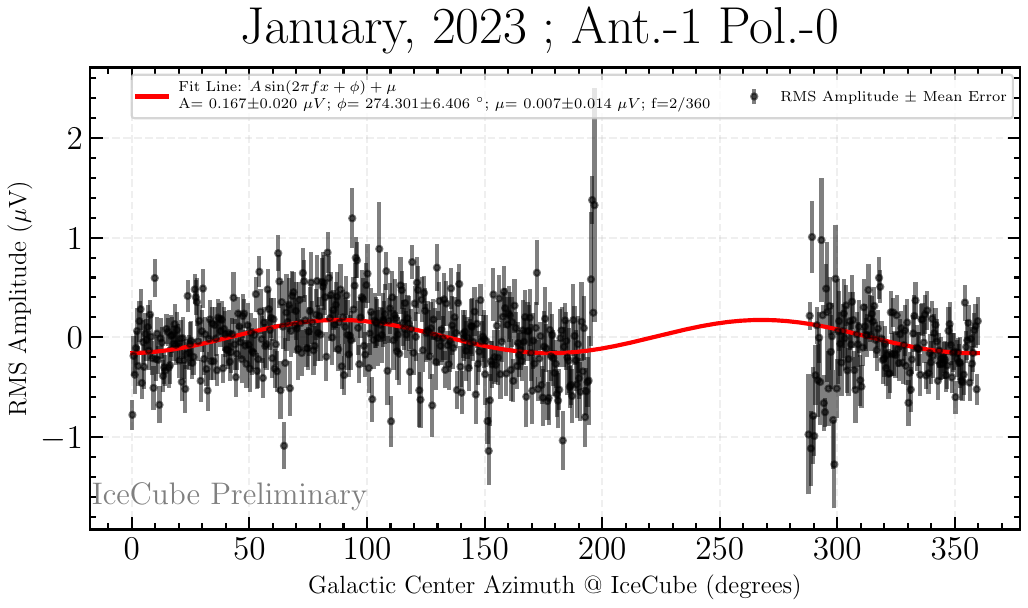}
    \end{minipage}\hfill
    \begin{minipage}{0.42\textwidth}
      \captionof{figure}{%
        RMS variation as a function of galactic center azimuth for a full month, revealing a sinusoidal pattern. The phase of this sinusoid indicates the azimuthal orientation of galactic center maxima, allowing determination of antenna arm alignment. Gaps occur because measurements aren't done throughout the day.
      }\label{fig:aziVar}
    \end{minipage}%
  }
\end{figure}

In order to obtain an unbiased dataset, software triggered events (referred to as fixed-rate trigger or FRT events) at regular intervals are utilized. The waveforms are sampled at a rate of 1000 MSps for the dataset we are using. However, the obtained waveforms in the different antenna polarizations have to be pre-processed. To perform the analysis, artifacts such as waveform spikes, RFI, and other noise sources (e.g., anthropogenic or electronic) must be addressed. This involves correcting or removing heavily contaminated events and cleaning the data by excluding outlier events. The following text details the implemented pipeline. The steps are also outlined in \autoref{fig:flowchart}. The steps are:
\begin{figure}
\centering
\includegraphics[width=1\linewidth]{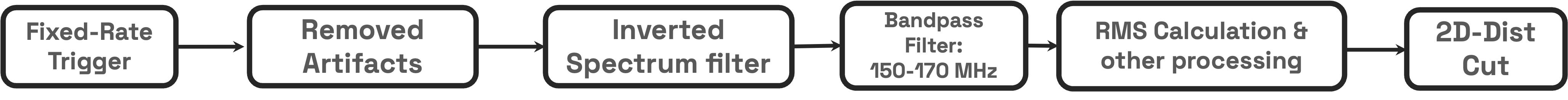}
\caption{Flowchart illustrating the pre-processing pipeline applied to waveforms prior to antenna orientation estimation. The steps include fixed-rate trigger event selection, artifact/spike removal (targeting anthropogenic and electronic noise), and spectral cleaning via an Inverted Spectrum Filter followed by a band-pass filter to suppress narrowband interference and isolate a specific galactic emission band. This is followed by galactic-noise strength estimation and outlier removal. For details check \autoref{sec:method}.}
\label{fig:flowchart}
\end{figure}
\begin{enumerate}
    \item \textbf{Removed Artifacts :} The waveforms obtained with the antennas are known to have multiple spikes. These can arise either because of bit-flips, or some transient noise-sources. In order to identify and remove such spikes we measure in a mode where four copies of the same waveform (per polarization for each antenna) are measured concurrently. A bin-wise comparison among the versions allows us to identify and correct a large amount of such spikes \cite{TurcotteTardif2023_1000160782}. To identify and remove the rest of the spikes a dynamic method was employed. The spike identification employs three different methods. These include identifying bins in the waveforms where the amplitude rises significantly higher than the rest of the waveforms, identifying sudden jumps and unsupervised machine-learning methods (isolation-forest). If the total number of unique spikes exceeds five per polarization then the event is removed. With the remaining events, for each polarization we first try with bit-flip (without changing the sign) in the most significant bit at the spike location. If the bit-flipped amplitude at the spike location is within five mean absolute deviation of the waveform bin amplitudes, then the amplitude at the spike location is assigned the bit-flipped value. Otherwise, the amplitude is assigned as the mean of the neighbors (one on both sides). If the spike is at the end (last-bin) of the waveform, then the mean is taken of the two bins on the left or right to it. During this procedure, it is ensured that the mean is taken of only those bins which are not identified as spike bins. Two example waveforms where this method has been employed are shown in \autoref{fig:waveformCleaning}. The top panels show the waveforms before any artifact removal. The vertical red-dashed lines show identified spike locations. The bottom panels shown the same waveforms after artifact removal. In each of the top panels, at least one of the identified spike location is clearly a spike (indicated by significant rise in amplitude). However, for the other locations making such a conclusion is difficult. The bottom panel shows that the corrected signal cleans the obvious spikes very well and has minimal information-loss in the other cases. Although more in-depth investigation is needed to pin-point the reason for the spikes, the current methodology prevents any sudden jumps in the waveforms.  Ongoing work tries to improve the identification and cleaning process further.\label{point:artifact}       
    \begin{figure}
    \centering
    \includegraphics[width=.8\linewidth]{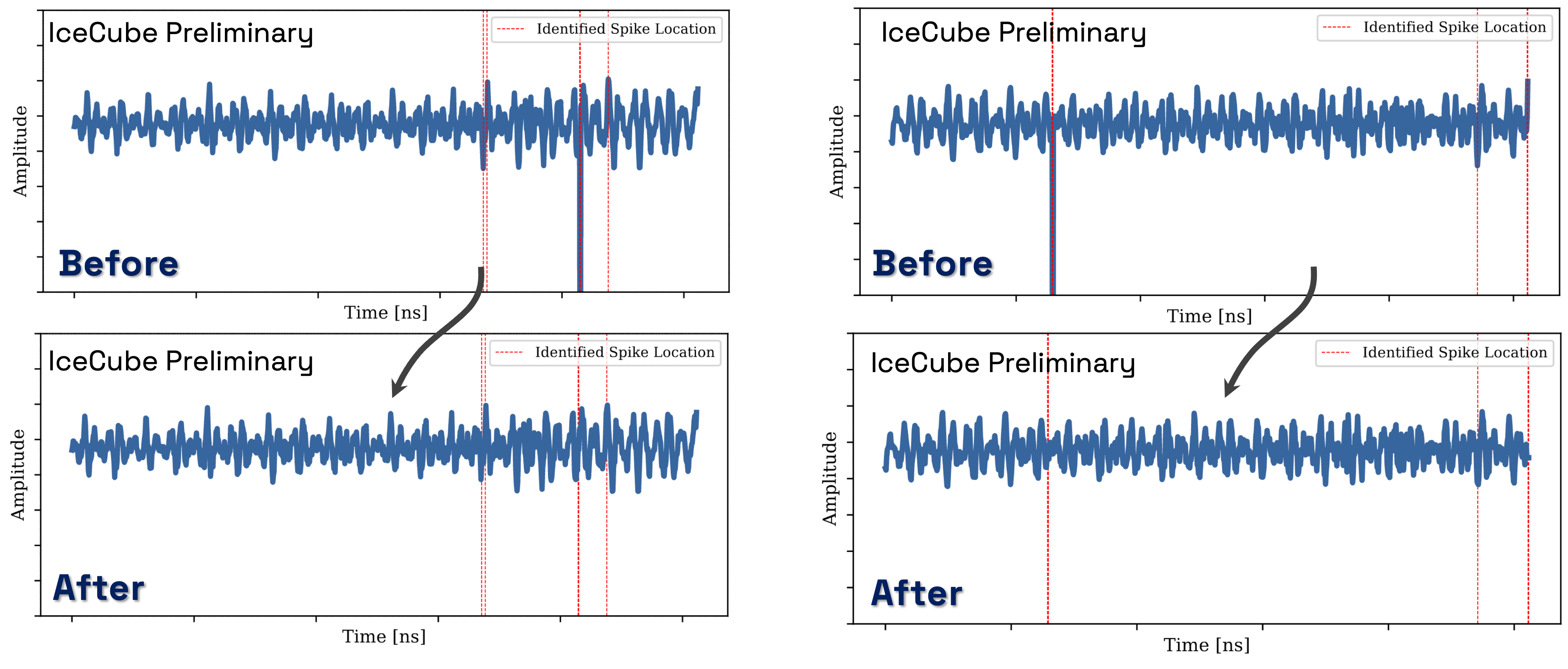}
    \caption{Two examples waveforms showing affect of artifact removal. For details check \autoref{point:artifact} (\autoref{sec:method}).}
    \label{fig:waveformCleaning}
    \end{figure}
    \item \textbf{Inverted Spectrum Filter and Bandpass Filter :} The previous step cleaned artifacts in the time-domain. The next task is to clean narrow-band spikes in the frequency domain because the Galactic noise is broad band. This involves two steps, Inverted Spectrum Filter and Band-Pass Filter, which include assessment of the waveforms in frequency spectrum and removing narrow-band spikes in the frequency spectrum (likely emerging from anthropogenic activity like communication channels and transient) and selecting a frequency band where the galactic radio emission strength is strong enough to be used for this analysis. To identify the spikes in frequency spectra for various polarizations a dynamic method to identify the spikes was developed \cite{Coleman2021}. The average frequency spectra of each month are first determined separately (shown in blue in \autoref{fig:freqSpectrum}). This is then utilized to calculate the median spectra by taking a sliding window of 31 MHz (green line labeled as "Median Spectrum" in \autoref{fig:freqSpectrum}). To identify the spikes the median spectra is divided by the measured spectrum (ratio in lower panel). The ratio is then utilized to get Filter Spectra by multiplication of the ratio with the measured spectra. This is represented as "Filter Applied x 1" in \autoref{fig:freqSpectrum} (in yellow color; also, offset for easier visibility). This filter can also be applied twice.  The galactic emission strength expectation is labeled as "Cane Model" \cite{Cane1979}. The 150-170 MHz frequency band was selected to mitigate affect of time and frequency domain artifacts/spikes in subsequent analyses, while ensuring that the galactic signal is also measurable.\label{point:spectrum}
    
    \begin{figure}[t]
      \centering
      \makebox[\textwidth][c]{%
        \begin{minipage}{0.7\textwidth}
          \includegraphics[width=\linewidth]{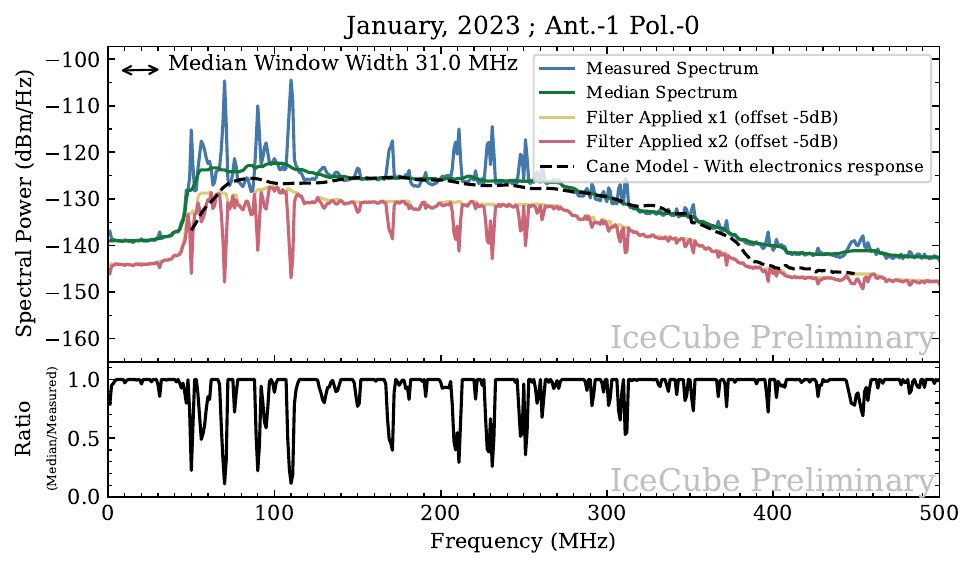}
        \end{minipage}\hfill
        \begin{minipage}{0.22\textwidth}
          \captionof{figure}{%
            Frequency Spectra for Antenna 1 - Polarization 0 for January, 2023. For details check \autoref{point:spectrum} (\autoref{sec:method}).
          }\label{fig:freqSpectrum}
        \end{minipage}%
      }
    \end{figure}
    \item \textbf{RMS Calculation :} The previous steps provide us with a subset of events and their corresponding waveforms with the majority of artifacts removed. To determine the galactic background strength from the recorded waveforms, we isolate a segment of each waveform that is as free as possible from air‐shower generated pulses or residual transient noise. Inspired by work done in Tunka-Rex \cite{tunkaRMS}, a moving window of size N = 120 samples (120 ns) is applied across the waveform and the RMS is calculated in each window and the minima of all the RMS values is assigned as the final RMS. This will be referred to as "RMS amplitude" throughout the text. The RMS in a window (k) is calculated using:   
     $\mathrm{RMS}_k = \sqrt{\frac{1}{N}\sum_{i=1}^{N}A_{k+i-1}^2}
   $ ( where A$_{k}$ is the amplitude at bin k). This allows to quantify the galactic signal strength per waveform and will be utilized later for antenna orientation estimation. Another quantity which will be utilized later is the variation in RMS. This is obtained by splitting the waveform into sub-traces (each of length 120 ns). RMS is calculated in each sub-trace. The 10 sub-traces with smallest RMS value are selected and the standard deviation of the corresponding RMS is calculated. This is referred to as "RMS Spread" throughout the text.  
    \item \textbf{2D RMS Cut :} After the initial cleaning process, a substantial number still exhibits residual artifacts beyond acceptable levels. This affects our orientation estimate. For now the most efficient method to remove such events (for this analysis) is to look at distribution of the RMS amplitudes and the RMS spread in waveforms. This is shown in \autoref{fig:2dCut}. Since we do not expect the galactic signal strength to change much throughout the year, we apply a cut such that the antenna polarization for the month with minimum number of outliers has minimal events being removed. This cut is kept the same for all polarizations and also kept constant throughout the year. For our case, in \autoref{fig:2dCut}, Antenna 1 - Polarization 0, October is the cleanest set and hence the cuts are visually selected to remove minimum events from this set. The removed events are represented by the shaded areas. 
    \begin{figure}
    \centering
    \includegraphics[width=.9\linewidth]{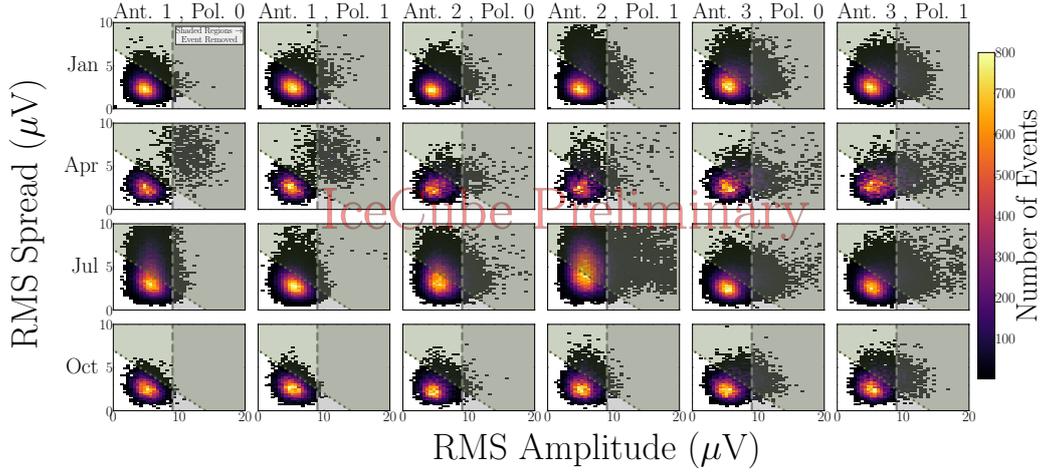}
    \caption{The 2D distribution of RMS values and their spread used to identify and remove noisy events that persist even after initial cleaning. The shaded regions represent the applied cuts, optimized to retain clean events while minimizing loss in months with the fewest outliers (e.g.,October for Antenna 1 – Polarization 0). For details check \autoref{sec:method}.}
    \label{fig:2dCut}
    \end{figure}
\end{enumerate}
After the initial cleaning and the 2D cut, we have a set of events where we have sufficiently reliable RMS distributions which can be used to look for time variation. This is shown in \autoref{fig:timeVar} for few days in January, 2023. Some oscillation are already visible at this stage in the RMS values which is in line with expectations from \autoref{sec:intro} since the azimuth of the GC makes a complete rotation around the antennas during a sidereal day.
To estimate the orientation from this information, we utilize the data for a month and plot RMS variation as a function of GC azimuth. This is shown in \autoref{fig:aziVar}. A sinusoidal is fit to this distribution and we obtain the GC azimuth where we see the maxima in the distribution. We now simply have to subtract/add 90 degrees to it to get antenna arm orientation. This is repeated separately for each polarization and for each month. The gaps in the time and azimuth variation plots simply arise from the fact that the measurements are not done throughout the entire day. The monthly orientation estimates using the galactic signal and a comparison with GPS estimates are discussed in \autoref{sec:results}.
\section{Results}\label{sec:results}
\begin{figure}
\centering
\includegraphics[width=.95\linewidth]{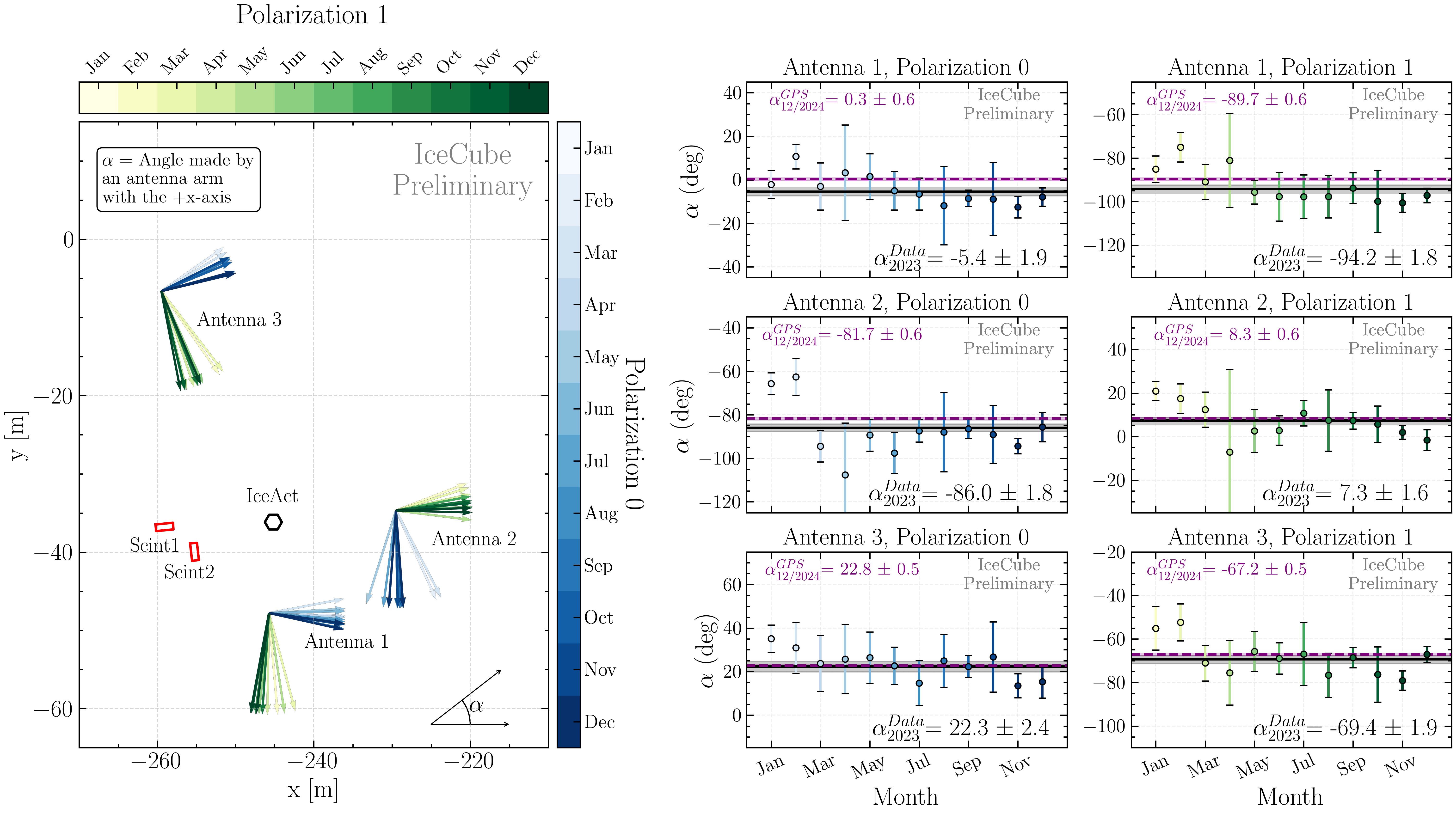}
\caption{Estimated orientation directions using the methodology outlined \autoref{sec:method}. Each rectangular subplot corresponds to a unique antenna and polarization. Monthly orientation estimates are shown with error bars, which are subsequently fit to obtain a yearly estimate. GPS-derived orientation is also indicated for reference. The left panel is an illustration of the mean orientation direction for each month for the three antennas. Other detectors, namely scintillators (labeled as Scint 1 and 2) and IceAct (an imaging air-cherenkov telescope)  are shown for reference.}
\label{fig:antennaOrient}
\end{figure}

The monthly antenna orientation estimates for each polarization using the methodology described in \autoref{sec:method} are shown in \autoref{fig:antennaOrient}. The left panel indicates the mean direction estimate for the two polarizations corresponding to the three antennas in the prototype station. In the estimated orientations, it is evident that apart from a few specific polarizations observed in certain months, the antenna polarizations are generally perpendicular to each other, as expected. The deviation in others might be because of anthropogenic and electronic noise. Polarization 0 (same for Polarization 1) for Antennas 1 and 3 points in a similar direction, but it varies for Antenna 2 due to a swap of polarization cables. Scintillators utilized to trigger the antennas (for EAS analysis) are also shown, labeled as Scint 1 and 2. 
\par 
The rectangular panels on the left in \autoref{fig:antennaOrient} also present the estimated uncertainty on each estimated antenna orientation. Each row corresponds to the two polarization in an antenna. A fit is performed to obtain yearly estimate of orientation for each antenna. The yearly fit with its uncertainty is represented as $\alpha_{2023}^{Data}$ in each of the rectangular panels. The expected perpendicular nature of the two polarizations in an antenna is clearly visible from the yearly fit values. The panels also show GPS measurements of antenna orientation. These are labeled as $\alpha_{12/2024}^{GPS}$ (done in December, 2024). The GPS orientation estimate is performed by measuring GPS coordinates of four known points on the antenna arms. As can be seen the GPS and data-driven estimates match very well for Antenna 3 (for both polarization). The slight tensions (in one or both polarization) which exist for the others are still within two standard deviations. Hence, this work presents a potential for the data-driven technique to potentially augment the GPS measurements.

\section{Summary and Outlook}\label{sec:outllok}
The work estimated antenna orientations at the prototype station surface array enhancement of IceCube Observatory by quantifying the galactic signal strength measurements and looking for their modulation as a function of the azimuth of Galactic center. The final obtained orientation measurements from data are in good agreement with GPS measurements on yearly basis. The agreement with GPS measurements indicates that this method is a promising approach for determining antenna orientation. Ongoing work to improve the various cleaning and filtering methods may potentially allow for a more frequent estimate of antenna orientations using galactic noise.     
\bibliographystyle{ICRC}
\bibliography{references}

\clearpage

\input{authorlist_IC.tex}
\end{document}

%% file: authorlist_IC.tex
\section*{Full Author List: IceCube Collaboration}

\scriptsize
\noindent
R. Abbasi$^{16}$,
M. Ackermann$^{63}$,
J. Adams$^{17}$,
S. K. Agarwalla$^{39,\: {\rm a}}$,
J. A. Aguilar$^{10}$,
M. Ahlers$^{21}$,
J.M. Alameddine$^{22}$,
S. Ali$^{35}$,
N. M. Amin$^{43}$,
K. Andeen$^{41}$
% ...
\\
\\
% ...
$^{10}$ Universit{\'e} Libre de Bruxelles, Science Faculty CP230, B-1050 Brussels, Belgium \\
% ...
$^{16}$ Department of Physics, Loyola University Chicago, Chicago, IL 60660, USA \\
$^{17}$ Dept. of Physics and Astronomy, University of Canterbury, Private Bag 4800, Christchurch, New Zealand \\
% ...
$^{21}$ Niels Bohr Institute, University of Copenhagen, DK-2100 Copenhagen, Denmark \\
$^{22}$ Dept. of Physics, TU Dortmund University, D-44221 Dortmund, Germany \\
% ...
$^{35}$ Dept. of Physics and Astronomy, University of Kansas, Lawrence, KS 66045, USA \\
% ...
$^{39}$ Dept. of Physics and Wisconsin IceCube Particle Astrophysics Center, University of Wisconsin{\textemdash}Madison, Madison, WI 53706, USA \\
% ...
$^{41}$ Department of Physics, Marquette University, Milwaukee, WI 53201, USA \\
% ...
$^{43}$ Bartol Research Institute and Dept. of Physics and Astronomy, University of Delaware, Newark, DE 19716, USA \\
% ...
$^{63}$ Deutsches Elektronen-Synchrotron DESY, Platanenallee 6, D-15738 Zeuthen, Germany \\
$^{\rm a}$ also at Institute of Physics, Sachivalaya Marg, Sainik School Post, Bhubaneswar 751005, India \\
% ...

\subsection*{Acknowledgments}

\noindent
The authors recognize the foundational contributions of Sara Nathalia Reina Torres and Roxanne Turcotte-Tardif, which were instrumental in initiating this work. The authors gratefully acknowledge the support from the following agencies and institutions:
USA {\textendash} U.S. National Science Foundation-Office of Polar Programs,
U.S. National Science Foundation-Physics Division,
U.S. National Science Foundation-EPSCoR,
U.S. National Science Foundation-Office of Advanced Cyberinfrastructure,
Wisconsin Alumni Research Foundation,
Center for High Throughput Computing (CHTC) at the University of Wisconsin{\textendash}Madison,
Open Science Grid (OSG),
Partnership to Advance Throughput Computing (PATh),
Advanced Cyberinfrastructure Coordination Ecosystem: Services {\&} Support (ACCESS),
Frontera and Ranch computing project at the Texas Advanced Computing Center,
U.S. Department of Energy-National Energy Research Scientific Computing Center,
Particle astrophysics research computing center at the University of Maryland,
Institute for Cyber-Enabled Research at Michigan State University,
Astroparticle physics computational facility at Marquette University,
NVIDIA Corporation,
and Google Cloud Platform;
Belgium {\textendash} Funds for Scientific Research (FRS-FNRS and FWO),
FWO Odysseus and Big Science programmes,
and Belgian Federal Science Policy Office (Belspo);
Germany {\textendash} Bundesministerium f{\"u}r Forschung, Technologie und Raumfahrt (BMFTR),
Deutsche Forschungsgemeinschaft (DFG),
Helmholtz Alliance for Astroparticle Physics (HAP),
Initiative and Networking Fund of the Helmholtz Association,
Deutsches Elektronen Synchrotron (DESY),
and High Performance Computing cluster of the RWTH Aachen;
Sweden {\textendash} Swedish Research Council,
Swedish Polar Research Secretariat,
Swedish National Infrastructure for Computing (SNIC),
and Knut and Alice Wallenberg Foundation;
European Union {\textendash} EGI Advanced Computing for research;
Australia {\textendash} Australian Research Council;
Canada {\textendash} Natural Sciences and Engineering Research Council of Canada,
Calcul Qu{\'e}bec, Compute Ontario, Canada Foundation for Innovation, WestGrid, and Digital Research Alliance of Canada;
Denmark {\textendash} Villum Fonden, Carlsberg Foundation, and European Commission;
New Zealand {\textendash} Marsden Fund;
Japan {\textendash} Japan Society for Promotion of Science (JSPS)
and Institute for Global Prominent Research (IGPR) of Chiba University;
Korea {\textendash} National Research Foundation of Korea (NRF);
Switzerland {\textendash} Swiss National Science Foundation (SNSF).